\documentclass[reprint,superscriptaddress,showpacs,aps,prb]{revtex4-2}

\usepackage[utf8]{inputenc}

\usepackage{graphicx}
\graphicspath{{.}{figures/}}

\usepackage{amsmath}
\usepackage{amssymb}
\usepackage{xfrac}

\usepackage{placeins}
\usepackage{siunitx}
\usepackage{enumitem}
\usepackage{xspace}
\usepackage{etoolbox}

\usepackage{pdfpages}  
\usepackage{pgffor}    
\makeatletter           
\AtBeginDocument{\let\LS@rot\@undefined}
\makeatother

\usepackage{xcolor}
\usepackage{hyperref}
\hypersetup{
    colorlinks,
    linkcolor={red!50!black},
    citecolor={blue!50!black},
    urlcolor={blue!80!black}
}

\usepackage{microtype}
\newcommand{\vect}[1]{{\boldsymbol{#1}}}

\newcommand{\inplane}[0]{\mbox{in-plane}\xspace}

\newcommand{\figautoref}[2][]{%
  \hyperref[#2]{%
    Fig.~\ref{#2}%
    \ifstrempty{#1}{}{\,#1}%
  }%
}

\newcommand{\figsautoref}[2][]{%
  \hyperref[#2]{%
    Figs.~\ref{#2}%
    \ifstrempty{#1}{}{\,#1}%
  }%
}

\DeclareSIUnit{\bigmicron}{\text{\Large\upmu}m}

\ExplSyntaxOn

\NewDocumentCommand{\eqrefstwo}{mm}{
  Equations~\hyperref[#1]{\ref*{#1}}\,--\,\hyperref[#2]{\ref*{#2}}\xspace
}

\ExplSyntaxOff

\ExplSyntaxOn

\NewDocumentCommand{\eqrefs}{m}{
  \mbox{\eqrefseries_process:n { #1 }}\xspace
}

\seq_new:N \l__eqrefseries_seq

\cs_new_protected:Npn \eqrefseries_process:n #1
 {
  \seq_set_split:Nnn \l__eqrefseries_seq { , } { #1 }

  \int_compare:nNnTF { \seq_count:N \l__eqrefseries_seq } = {1}
   {
    Equation~\hyperref[\seq_item:Nn \l__eqrefseries_seq {1}]{\ref*{\seq_item:Nn \l__eqrefseries_seq {1}}}%
   }
   {
    Equations~\eqrefseries_print_list:
   }
 }

\cs_new_protected:Npn \eqrefseries_print_list:
 {
  \int_zero_new:N \l_tmpa_int
  \seq_map_inline:Nn \l__eqrefseries_seq
   {
    \int_incr:N \l_tmpa_int
    \int_compare:nNnTF { \l_tmpa_int } = {1}
     {
      \hyperref[##1]{\ref*{##1}}\xspace%
     }
     {
      \int_compare:nNnTF { \l_tmpa_int } = {\seq_count:N \l__eqrefseries_seq}
       {
        ~and~\hyperref[##1]{\ref*{##1}}%
       }
       {
        ,~\hyperref[##1]{\ref*{##1}}\xspace%
       }
     }
   }
 }

\ExplSyntaxOff

\ExplSyntaxOn

\NewDocumentCommand{\figrefstwo}{mm}{
  Figures~\hyperref[#1]{\ref*{#1}}\,--\,\hyperref[#2]{\ref*{#2}}\xspace
}

\NewDocumentCommand{\figrefs}{m}{
  \mbox{\figrefseries_process:n { #1 }}\xspace
}

\seq_new:N \l__figrefseries_seq

\cs_new_protected:Npn \figrefseries_process:n #1
 {
  \seq_set_split:Nnn \l__figrefseries_seq { , } { #1 }

  \int_compare:nNnTF { \seq_count:N \l__figrefseries_seq } = {1}
   {
    Figure~\hyperref[\seq_item:Nn \l__figrefseries_seq {1}]{\ref*{\seq_item:Nn \l__figrefseries_seq {1}}}%
   }
   {
    Figures~\figrefseries_print_list:
   }
 }

\cs_new_protected:Npn \figrefseries_print_list:
 {
  \int_zero_new:N \l_tmpa_int
  \seq_map_inline:Nn \l__figrefseries_seq
   {
    \int_incr:N \l_tmpa_int
    \int_compare:nNnTF { \l_tmpa_int } = {1}
     {
      \hyperref[##1]{\ref*{##1}}\xspace%
     }
     {
      \int_compare:nNnTF { \l_tmpa_int } = {\seq_count:N \l__figrefseries_seq}
       {
        ~and~\hyperref[##1]{\ref*{##1}}%
       }
       {
        ,~\hyperref[##1]{\ref*{##1}}\xspace%
       }
     }
   }
 }

\ExplSyntaxOff

\newcommand{\tern}[6]{{#1$_{#2}$#3$_{#4}$#5$_{#6}$}\xspace}  						

\newcommand{\MnPtSn}[0]{\tern{Mn}{1.4}{Pt}{}{Sn}{}}

\newcommand{\MPI}{Max Planck Institute for
Chemical Physics of Solids, 01187 Dresden, Germany.}
\newcommand{\DCN}{Dresden Center for Nanoanalysis, cfaed, Technical University Dresden, 01069 Dresden, Germany.}

\newcommand{\IFMP}{Institute for Solid State and Materials Physics, Technical University of Dresden, 01062 Dresden, Germany.}
\newcommand{\DiamondLight}{Diamond Light Source, Harwell Science and Innovation Campus, Didcot, OX11~0DE, United Kingdom}
\newcommand{\Clarendon}{Department of Physics, Clarendon Laboratory, University of Oxford, Oxford, OX1~3PU, United Kingdom}

\newcommand{\Bessy}{Helmholtz-Zentrum Berlin für Materialien und Energie, D-12489 Berlin, Germany}

\newcommand{\UNIDue}{Faculty of Physics and Center for Nanointegration Duisburg-Essen (CENIDE), University of Duisburg-Essen, 47057 Duisburg, Germany}
\newcommand{\ctqmat}{Würzburg-Dresden Cluster of Excellence ct.qmat, Technische Universität Dresden, 01062 Dresden, Germany}
\newcommand{\augsburg}{Experimental Physics VI, Center for Electronic Correlations and Magnetism, University of Augsburg, 86159 Augsburg, Germany}

\begin{document}

\graphicspath{{.}{figures/}}

\title{Quasi-1D Spin Textures: From Chiral Soliton Lattice to Fan State}





\author{M. Winter}
\affiliation{\MPI}
\affiliation{\DCN}
\affiliation{\augsburg}
\affiliation{\DiamondLight}

\author{A. Pignedoli}
\affiliation{\UNIDue}

\author{A. S. Sukhanov}
\affiliation{\augsburg}


\author{M. Azhar}
\affiliation{\UNIDue}

\author{A. Tahn}
\affiliation{\DCN}

\author{B. Achinuq}
\affiliation{\Clarendon}

\author{\mbox{J. R. Bollard}}
\affiliation{\Clarendon}
\affiliation{\DiamondLight}

\author{V. Ukleev}
\affiliation{\Bessy}

\author{C. Luo}
\affiliation{\Bessy}

\author{F. Radu}
\affiliation{\Bessy}

\author{S. Wintz}
\affiliation{\Bessy}

\author{M. Weigand}
\affiliation{\Bessy}

\author{A. Mistonov}
\affiliation{\IFMP}

\author{P. Vir}
\affiliation{\MPI}

\author{J. Geck}
\affiliation{\IFMP}
\affiliation{\ctqmat}

\author{\mbox{C. Felser}}
\affiliation{\MPI}
\affiliation{\ctqmat}

\author{G. van der Laan}
\affiliation{\DiamondLight}

\author{T. Hesjedal}
\affiliation{\Clarendon}
\affiliation{\DiamondLight}

\author{K. Everschor-Sitte}
\affiliation{\UNIDue}

\author{B. Rellinghaus}
\affiliation{\DCN}

\author{M. C. Rahn}
\affiliation{\augsburg}
\date{June 19, 2026}

\begin{abstract}
In most helimagnets, an applied magnetic field aligns the propagation direction of a helical spin texture with the field, resulting in a conical state and obscuring the unwinding process. Here, we access a complementary regime in the anisotropic chiral magnet Mn$_{1.4}$PtSn, where crystal symmetry constrains the propagation direction of the spin modulation. Using resonant elastic X-ray scattering in a vector magnet, we track the evolution of quasi-one-dimensional spin textures that propagate along a chiral crystallographic axis while the magnetic field is applied perpendicular to this direction. Together with micromagnetic simulations, our measurements reveal a transformation from the zero-field $\pi$-chiral soliton lattice into a fan-like state. In this state, the propagation direction remains transverse to the applied field, while the spins oscillate about the field direction. During magnetization, the modulation length decreases continuously with the field and approaches the field-polarized state. Simulations indicate that magnetostatic interactions in finite samples play a key role in stabilizing this behavior. Our results provide evidence for a fan-like regime in a chiral magnet and highlight how field orientation can be used to control one-dimensional spin textures.

\end{abstract}

\maketitle

\section{Introduction}

Chiral magnetic materials have emerged as a fertile ground for exploring topologically nontrivial spin textures, such as skyrmions, chiral soliton lattices (CSLs), and related modulated states~\cite{Nagaosa2013TopologicalSkyrmions,Fert2013,koraltan20262026skyrmionicsroadmap}. These textures are stabilized by competing interactions—exchange, Dzyaloshinskii–Moriya (DMI), Zeeman, and magnetocrystalline anisotropy—and exhibit unique transport and dynamical properties that make them promising for spintronic and magnonic applications. Their ability to encode information in topological invariants, combined with efficient manipulation by electric currents or magnetic fields, has inspired concepts for racetrack memories, logic devices, and neuromorphic architectures~\cite{Fert2013,Zhang2021}. While two-dimensional textures such as skyrmions have been widely studied, one-dimensional modulations oﬀer in principle a simpler setting in which field-driven transformations, such as soliton unwinding, can be tracked directly. Understanding how these 1D textures evolve under external stimuli is, therefore, crucial for both fundamental physics and device engineering.

The present work provides new insights into one-dimensional spin textures in the Heusler compound Mn$_{1.4}$PtSn with $D_{{2d}}$ point symmetry (space group $I\bar{4}2d$). To date, studies of magnetic textures in this material have primarily focused on real-space measurements on thin samples with the crystallographic $c$ axis oriented out-of-plane, which allows the observation of antiskyrmions (aSks) and related multi-$\vect{q}$ states~\cite{Nayak2017,Ma2020,Peng2020,Peng2022,Winter2022,ZunigaCespedes2021,Jena2020a}. Lorentz transmission electron microscopy (LTEM), which provides a magnetic field perpendicular to the sample platelet, is highly suited to explore these phenomena~\cite{Nayak2017,Peng2020,Jena2020a,Jena2022ObservationMaterial,Jena2024}. Owing to the material's easy-axis anisotropy, even slight tilts of the magnetic field away from the $c$ axis induce pronounced changes in the spin configuration, including the formation of nontopological bubbles (NTbs)~\cite{Peng2020,Jena2020a,Jena2022ObservationMaterial,Jena2024}. However, the experimental limitations of electron microscopy imply that the in-plane and out-of-plane response of highly anisotropic materials cannot be separated. Despite the evident sensitivity of its magnetic textures to in-plane fields, this regime has therefore remained largely unexplored in  Mn$_{1.4}$PtSn~\cite{Sukhanov2022,Ma2020}.

In recent work, the ground state of \MnPtSn was identified as a $\pi$-CSL~\cite{Winter2025}. As illustrated in \figautoref[(a,b)]{fig:FAN_introduction}, this corresponds to a periodic array of $\pi$ domain walls mediating $180^\circ$ rotations of the magnetization vector around the (in-plane) propagation axis. Under out-of-plane fields, neighboring $\pi$-solitons condense into $2\pi$ rotations and form a $2\pi$-CSL that evolves toward the spin-polarized state as the soliton density decreases.

\begin{figure*}[tp]
    \centering
    \includegraphics[width=\textwidth]{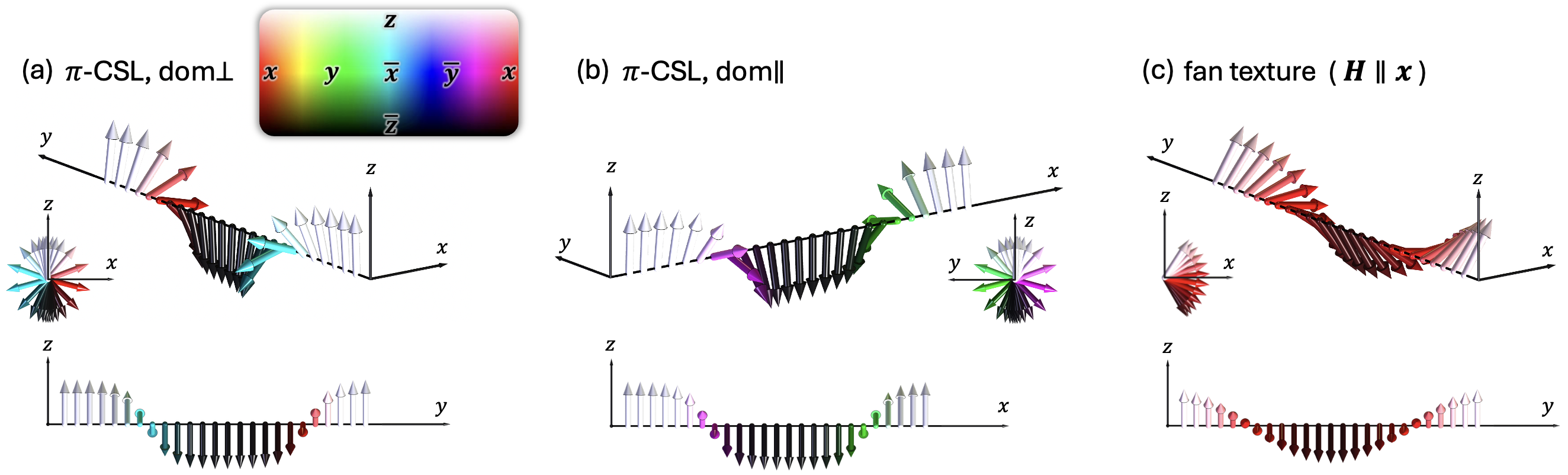}
    \caption{Representations of one-dimensional spin textures in Mn$_{1.4}$PtSn. Parallel and perpendicular domains (dom$_\parallel$, dom$_\perp$) are labeled assuming a magnetic field oriented along the $x$-axis. The characteristic rotation of the spins is illustrated by arrows in the \textit{hsl} color space (hue/saturation/lightness, see inset). (a) $\pi$-CSL domain propagating along the $y$-axis. The soliton lattice forms as a periodic array of $\pi$ Bloch domain walls. Considering the spin texture in two spatial dimensions, the magnetization field $\mathbf{m}(\mathbf{r})$ defines a mapping $\mathbb{R}^2 \to S^1$, characterized by a finite winding number $Q \neq 0$. (b) Corresponding $\pi$-CSL domain propagating along the $x$-axis, with opposite helicity. (c) Fan state observed in a field along the $x$-axis: spins oscillate symmetrically about the field direction without completing a full rotation ($Q = 0$).}
    \label{fig:FAN_introduction}
\end{figure*}

Building on this foundation, we here address the complementary case of in-plane fields applied perpendicular to the propagation axis. We show that in this geometry, the $\pi$-CSL does not condense into higher-order solitons but instead transforms into a distinct modulated texture, the fan state illustrated in \figautoref[(c)]{fig:FAN_introduction}. Considering the effective two dimensional magnetization as an $S^1$ order parameter, this corresponds to a topological transition: while the $\pi$-CSL carries a finite winding number ($Q \neq 0$), the fan is topologically trivial ($Q = 0$), with spins oscillating symmetrically about the field direction. Crucially, overcoming the net winding allows the continuous connection to the fully spin-polarized state.

Fan-type textures are well known in classical Yoshimori-type helimagnets without DMI~\cite{Izyumov1984,Yoshimori1959}, where competing exchange interactions stabilize helices of opposite chirality that combine under perpendicular fields into a fan~\cite{Ishikawa1969}. In DMI-active systems such as Mn$_{1.4}$PtSn, however, the fixed rotational sense penalizes the fan configuration, and its stability has remained uncertain. Although fan-like states have been predicted by simulations~\cite{Osorio2023} in monoaxial helimagnets, corresponding LTEM studies~\cite{Karna2021,Hall2022} could not definitively resolve them from chiral modulations due to the measurement geometry employed. Simulations suggest that fans can appear under external fields~\cite{Osorio2023} and persist metastably at reduced fields through disorder or thermal fluctuations~\cite{Shinhozaki2017}. Whether a fan state can be realized as a stable equilibrium phase in a DMI-active system---and what mechanisms enable its stabilization---remains an open question.

Here, we provide experimental evidence of a fan state in a chiral magnet. Using field-dependent REXS measurements combined with two specifically designed micromagnetic simulation approaches, we demonstrate that an in-plane magnetic field drives an irreversible transition from a mixed-domain $\pi$-CSL ground state to a single-domain fan configuration. The simulations reproduce the observed domain redistribution and the field dependence of the texture's wave number. Our results reveal a field-driven transition from a chiral soliton lattice to a fan-like state and highlight the crucial role of field orientation in shaping one-dimensional spin textures in Mn$_{1.4}$PtSn.

\section{Experimental results}

High-quality single crystals of Mn$_{1.4}$PtSn were grown by the self-flux method. As detailed in Ref.~\cite{Winter2025}, oriented crystallites were machined into lamellae with the \textit{c}-axis as the surface normal using focused ion beam (FIB) milling. To implement a resonant elastic X-ray scattering study in transmission geometry, a \qty{19}{\micro\metre}~$\times$~\qty{14.7}{\micro\metre} lamella was prepared with a thickness of approximately \qty{80}{\nano\metre} $\pm$ \qty{20}{\nano\metre}
on the order of the attenuation length at the Mn $L_3$ edge (637\,eV). This lamella was mounted on top of an \qty{8}{\micro\metre}~$\times$~\qty{8}{\micro\metre} square aperture that had been cut into a molybdenum disc. This disc (\qty{3}{\milli\metre} diameter, \qty{50}{\micro\metre} thickness) effectively absorbed the beam and served as a sample holder.

\begin{figure*}[tp]
    \centering
    \includegraphics[width=\textwidth]{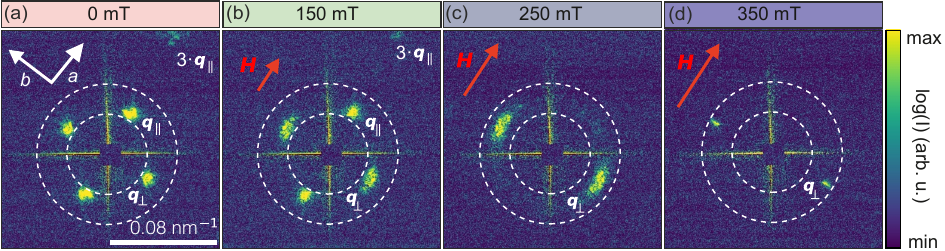}
    \caption{(a) Background-subtracted small-angle REXS patterns of Mn$_{1.4}$PtSn showing momentum transfers in the $a$--$b$ plane (all data at room temperature). Panels (b-d) illustrate characteristic changes in a magnetic field of up to 350\,mT applied along the $a$-axis, as indicated by red arrows. The mixed-domain $\pi$-CSL state persists at 150\,mT (b) before the domain with propagation vector $\vect{q}_\parallel$ along the field collapses (c), and a single-domain state with fewer imperfections, i.e., narrow $\vect{q}_\perp$ Bragg spots, emerges at high fields (d). The two dashed circles serve as a guide to the eye to indicate the range over which the first-order Bragg peaks evolve during the field sweep; this annular region was used for the quantitative analysis in \figautoref{fig_fieldsweep_0_to_sat_bragg_only_magnitude}. The data were obtained while ramping the field up. Scattered intensity is presented on a logarithmic color scale (cf.\ colorbar) with an additional median filter.}
    \label{fig:Results_B_fieldsweep_0_to_sat_exp}
\end{figure*}

REXS measurements under \inplane field protocols were conducted using the portable octupole magnet setup (POMS) at beamline I10, Diamond Light Source. A sequence of background-subtracted REXS patterns is shown in \figautoref{fig:Results_B_fieldsweep_0_to_sat_exp}. The data were obtained during a field sweep where the field, applied parallel to the $a$ axis, was incremented in steps of $\SI{10}{\milli\tesla}$ from zero to $\SI{450}{\milli\tesla}$. Additional details, including data obtained during the down-ramp, are provided in Sec.~S-I of the Supplemental Material.

\begin{figure}[tbp]
    \centering
    \includegraphics[width=0.48\textwidth]{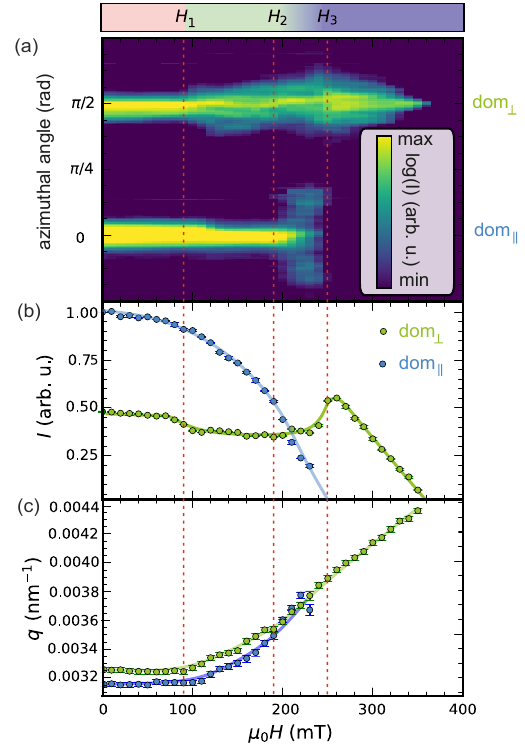}
    \caption{Quantitative analysis of the REXS Bragg peaks for dom$_\perp$ (green) and dom$_\parallel$ (blue) in increasing \inplane fields (cf.~\figautoref{fig:Results_B_fieldsweep_0_to_sat_exp}). Dashed red  lines indicate the characteristic fields and the colored band at the top mark the corresponding phase regions. (a)~Radially integrated intensity in one half of the scattering pattern within the annular $q$~range indicated in \figautoref{fig:Results_B_fieldsweep_0_to_sat_exp}. The intensity is shown on a logarithmic color scale as a function of the azimuthal angle and applied field. The two horizontal bands near $0$ and $\pi/2$ correspond to the Bragg peaks of dom$_\parallel$ and dom$_\perp$, respectively. Above $H_1$, the dom$_\perp$ peak broadens visibly; above $H_2$, the dom$_\parallel$ peak broadens and fades and vanishes at $H_3$. (b)~Bragg peak intensities $I_\perp$ and $I_\parallel$. $I_\parallel$ decreases monotonically, with a steep drop above $H_2$. $I_\perp$ shows a slight decrease at $H_1$, followed by a broad maximum near $H_3$ before declining toward saturation. (c)~The texture wave numbers $q_\perp$ and $q_\parallel$. Both remain on a constant plateau up to $H_1$ and increase above it; $q_\parallel$ is last resolved near $H_3$, while $q_\perp$ continues to rise up to $\sim\SI{350}{\milli\tesla}$.}
    \label{fig_fieldsweep_0_to_sat_bragg_only_magnitude}
\end{figure}

    The magnetic ground state of Mn$_{1.4}$PtSn consists of two orthogonal stripe-domains aligned with the crystallographic $a$ and $b$ axes~\cite{Nayak2017,Ma2020,Peng2020,ZunigaCespedes2021,Jena2020a}. As seen in \figautoref[(a)]{fig:Results_B_fieldsweep_0_to_sat_exp}, this results in a pseudo-fourfold REXS pattern, where either pair of Bragg peaks corresponds to one domain. The peaks labeled $\vect{q}_\parallel$ originate from the domain propagating along the applied in-plane field (dom$_{\parallel}$), while those labeled $\vect{q}_\perp$ correspond to the domain with propagation perpendicular to the field (dom$_{\perp}$). Due to the $D_{2d}$ symmetry, the $\pi$-CSL textures of either domain ($\pi$-CSL$_{\parallel}$, $\pi$-CSL$_{\perp}$) have opposite chirality~\cite{Nayak2017,Peng2020,Winter2025}. The presence of the third harmonic at $3\vect{q}_\parallel$ indicates a non-sinusoidal modulation characteristic of a $\pi$-CSL, as discussed in Ref.~\cite{Winter2025}.

As seen in \figautoref[(b)]{fig:Results_B_fieldsweep_0_to_sat_exp}, the pattern remains unchanged at $\SI{150}{\milli\tesla}$. At $\SI{250}{\milli\tesla}$, the $\vect{q}_\parallel$ peak disappears, marking the collapse of
the domain propagating along the field and a transfer of the phase volume to dom$_{\perp}$. The significant broadening of the Bragg peaks above this domain-re-population signals the proliferation of lattice imperfections [cf.~\figautoref[(c)]{fig:Results_B_fieldsweep_0_to_sat_exp}]. By $\SI{350}{\milli\tesla}$, the system has attained a coherent long-ranged modulation, recognizable by the resolution-limited Bragg reflections in \figautoref[(d)]{fig:Results_B_fieldsweep_0_to_sat_exp}. To guarantee full saturation, the field was increased to $\SI{450}{\milli\tesla}$. The down-sweep shown in the Supplemental Material (Fig.~S1) reveals that, upon reducing the field back to zero, the sample remains in a single-domain state: only the Bragg peak at $\vect{q}_\perp$ persists, demonstrating that
the collapse of dom$_{\parallel}$ is irreversible.

\begin{figure*}[tp]
    \centering
    \includegraphics[width=\textwidth]{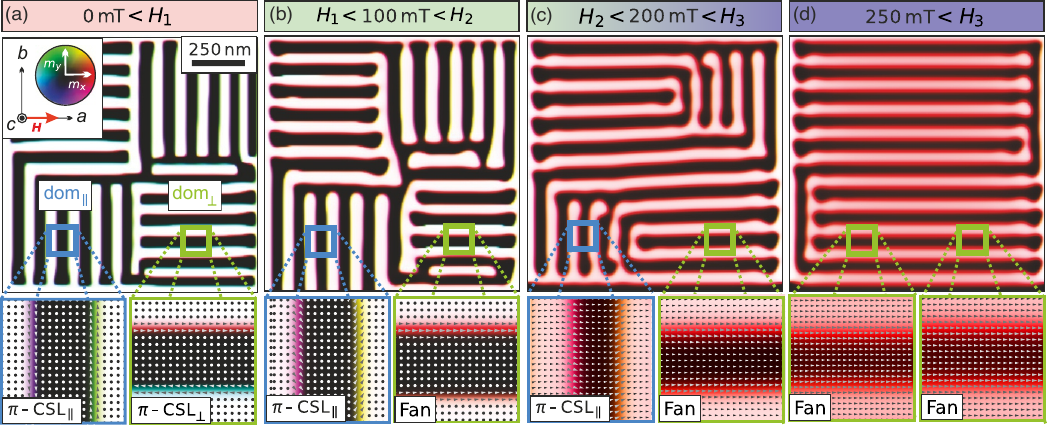}
    \caption{Micromagnetic simulations of spin textures in a $2\times2\times0.1\,\si{\micro\metre}^3$ slab of Mn$_{1.4}$PtSn in fields up to 250\,mT along the $a$-axis. As indicated in the inset, the spin orientations in the central ($z=50$\,nm) cross-sectional plane are rendered on an \textit{hsl} color coordinate: hue and saturation encode the in-plane magnetization, while lightness represents the out-of-plane component. (a) The ground state at zero field is initialized with right-handed chiral soliton lattice domains propagating along the $a$ axis (dom$_{\parallel}$), and left-handed domains perpendicular (dom$_{\perp}$). The green and blue boxes and corresponding bottom panels provide detailed views illustrating the type of spin rotation between up- (white) and down (black) stripes. A sequence of interfaces with opposite colors (e.g., purple/green) identifies a $\pi$-CSL texture. (b) At 100\,mT, i.e., above $H_1$, dom$_\parallel$ features a conically deformed (pink/yellow) $\pi$-CSL, while dom$_\perp$ is transformed to a fan texture (red/red). (c) At 200\,mT, i.e., above $H_2$, dom$_\perp$ has significantly gained in phase volume. (d) At 250\,mT, i.e., above $H_3$, the dom$_\parallel$ domain has been fully displaced by the fan structure, which continuously connects to the spin-polarized state.}
    \label{Fig:micromagnetic_neumann}
\end{figure*}

\figautoref{fig_fieldsweep_0_to_sat_bragg_only_magnitude} presents a detailed quantitative analysis of the REXS signal as a function of applied field. \figautoref[(a)]{fig_fieldsweep_0_to_sat_bragg_only_magnitude} displays the radially integrated intensity within the annular $q$~range indicated by the dashed circles in \figautoref{fig:Results_B_fieldsweep_0_to_sat_exp}, plotted against the azimuthal angle. Three characteristic field ranges of the magnetization process can be recognized at a glance: Up to $H_1 \approx \SI{90}{\milli\tesla}$, the scattering pattern is essentially unchanged. Above $H_1$, notable disorder sets in: the Bragg peak of dom$_\perp$ broadens, which signals the onset of structural rearrangement in this domain. Crossing $H_2 \approx \SI{190}{\milli\tesla}$, the Bragg peak at dom$_\parallel$ broadens azimuthally (or directionally) and its intensity decreases rapidly; it vanishes above $H_3 \approx \SI{240}{\milli\tesla}$, which marks the formation of the single-domain state. Beyond $H_3$, the broadening of the dom$_\perp$ peak progressively reduces until saturation is reached. To corroborate this interpretation of the magnetization process in real space, we performed scanning transmission X-ray microscopy (STXM) scans of a comparable sample at selected fields. For illustration, the resulting real-space images of the domain reorganization are presented in Sec.~S-II of the Supplemental Material.

As shown in \figautoref[(b)]{fig_fieldsweep_0_to_sat_bragg_only_magnitude}, the integrated intensities ($I_\perp$, $I_\parallel$) of the Bragg peaks characterize the domain-selective response to the critical fields. At low fields, the intensity ratio reveals an (arbitrary) domain population of approximately {3:2}. At $H_1$, a drop in $I_\perp$ is observed, while $I_\parallel$ remains unaffected. This asymmetry suggests that dom$_\perp$ undergoes a reduction in the out-of-plane magnetization component $m_z$. We return to this aspect in the discussion of a magnetization jump observed in micromagnetic simulations. Above $H_2$, $I_\parallel$ decreases rapidly, consistent with the collapse of dom$_\parallel$.

The field evolution of the wave numbers in \figautoref[(c)]{fig_fieldsweep_0_to_sat_bragg_only_magnitude} shows that up to $H_{\mathrm{1}}$ the modulation lengths of both domains remain constant with a value of $\sim 2\pi/q_{\parallel,\perp} = 150$--$160$\,nm (the slight discrepancy can likely be assigned to the rectangular aspect ratio or limited flatness of the sample). Above $H_{\mathrm{1}}$, the modulations gradually contract ($q_{\parallel,\perp}$ increase). The contraction of dom$_\parallel$ persists up to $H_3$, where all remaining intensity is transferred to dom$_\perp$. Beyond $H_{\mathrm{3}}$, the spin-polarization and contraction of this domain continue at a constant rate of $0.16\,$nm/mT, down to a modulation length of 115\,nm at 350\,mT, where the material enters the saturated state.

The familiar consequence of spin-polarizing conventional helimagnets is the formation of a conical structure, associated with a magnetization along the propagation vector, which aligns with the field direction. This scenario would indeed be plausible, except for the behavior above $H_{\mathrm{2}}$. As a conical $q_\parallel$-domain would be more favorable, the selection of $q_\perp$ around $H_{\mathrm{2}}\sim H_{\mathrm{3}}$ rules out the possibility that the field has induced a conical modulation. Candidates for 1D modulated structures favoring propagation perpendicular to the field would be an anharmonic helix or CSL where, along the direction of propagation, regions of largely field-aligned spins grow disproportionately~\cite{Brearton2023,Okamura2017,Ukleev2020}. However, this scenario would be associated with a stretched modulation length~\cite{Togawa2012,Izyumov1984,Dzyaloshinskii1964} and at odds with the characteristics of \figautoref[(c)]{fig_fieldsweep_0_to_sat_bragg_only_magnitude}.

\section{Micromagnetic simulations}

To interpret our experimental observations, we performed micromagnetic simulations modeling \MnPtSn\ under $D_{{2d}}$ point symmetry with the full micromagnetic Hamiltonian including exchange, anisotropic DMI, Zeeman, magnetocrystalline anisotropy, and magnetostatic contributions. The complete details are provided in the Supplemental Material, where Sec.~S-III\,A specifies the micromagnetic Hamiltonian and Sec.~S-III\,B lists the material parameters used in the simulations.

\subsection{\texorpdfstring{$\pi$}{pi}-CSL-to-fan transition and domain-redistribution in in-plane fields}

First, to investigate the irreversible domain collapse observed in REXS, we initialized a $2\times2\times0.1\,\si{\micro\metre}^3$ slab with Neumann (free) boundary conditions in a mixed-domain state consistent with the experimental ground state. To simulate the field sweep, this spin configuration was consecutively relaxed at a sequence of fields up to $\SI{450}{\milli\tesla}$ applied along the $x$ direction, corresponding to the crystallographic $a$ axis. While chiral surface twists at the sample boundary introduce energy barriers not fully captured in this geometry, the simulations nonetheless provide a qualitative account of the observed domain behavior. Full details of this finite-sample, open-boundary approach are given in Sec.~S-III\,C of the Supplemental Material.

The resulting magnetic textures are shown in \figautoref{Fig:micromagnetic_neumann} for four representative field strengths. Regions of interest (ROIs) representing the two domains (dom$_\parallel$, dom$_\perp$) are marked by blue and green boxes, respectively, and corresponding detailed views of the magnetization textures are shown in the bottom panels. The spin direction is visualized by the hue-saturation-lightness (\textit{hsl}) color coordinate. As illustrated in \figautoref{fig:FAN_introduction}, this indicates the sense of the spin-rotation by the color sequence of interfaces between down (black) and up (white) stripes. For instance, at zero field [cf.~\figautoref[(a)]{Fig:micromagnetic_neumann}], the rotations through ``opposite'' colors (purple/green, red/cyan) reveal that the two domains host $\pi$-CSLs of opposite chirality oriented perpendicular to one another: $\pi$-CSL$_\parallel$ with right-handed and $\pi$-CSL$_\perp$ with left-handed chirality.

\begin{figure}[tbp]
    \centering
    \includegraphics[width=0.48\textwidth]{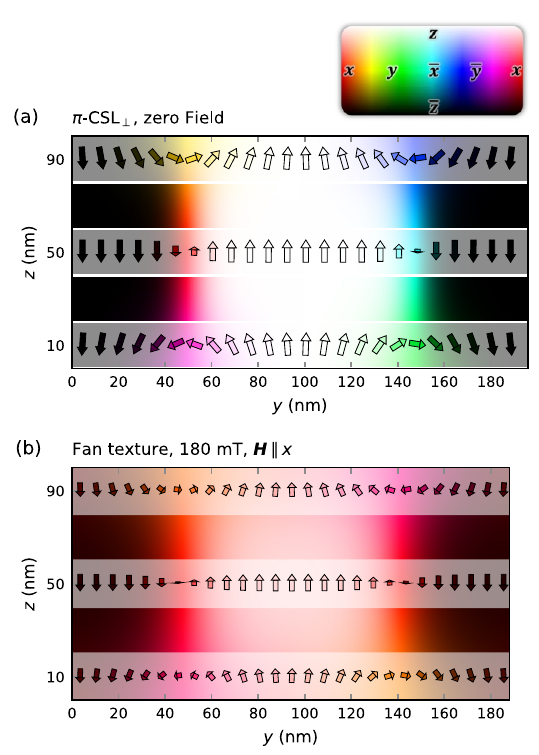}
    \caption{Micromagnetic simulations illustrating the effects of demagnetization fields on spin textures in a 100\,nm Mn$_{1.4}$PtSn lamella. The magnetization direction is drawn on the \textit{hsl} color scale, as in \figautoref{Fig:micromagnetic_neumann}. To model the finite sample size, periodic boundary conditions are applied in the in-plane directions ($x$,$y$), while the top and bottom layers of the slab ($z = \SI{10}{\nano\meter}$ and $\SI{90}{\nano\meter}$) are unconstrained. (a) In zero field, a model $\pi$-CSL texture is realized in the bulk ($z = \SI{50}{\nano\meter}$), with a Bloch-like rotation of the spins in a plane perpendicular to the direction of propagation ($y$). At the surfaces, flux-closure favors a distortion towards Néel-like solitons with sizable magnetization components along $y$. (b) With 180\,mT applied along the $x$-direction, a fan-texture of reduced modulation length is realized. The effects of flux-closure distortion remain observable at the surfaces.}
    \label{fig:micro_PBCs}
\end{figure}

\begin{figure}[tbp]  
    \centering
    \includegraphics[width=0.48\textwidth]{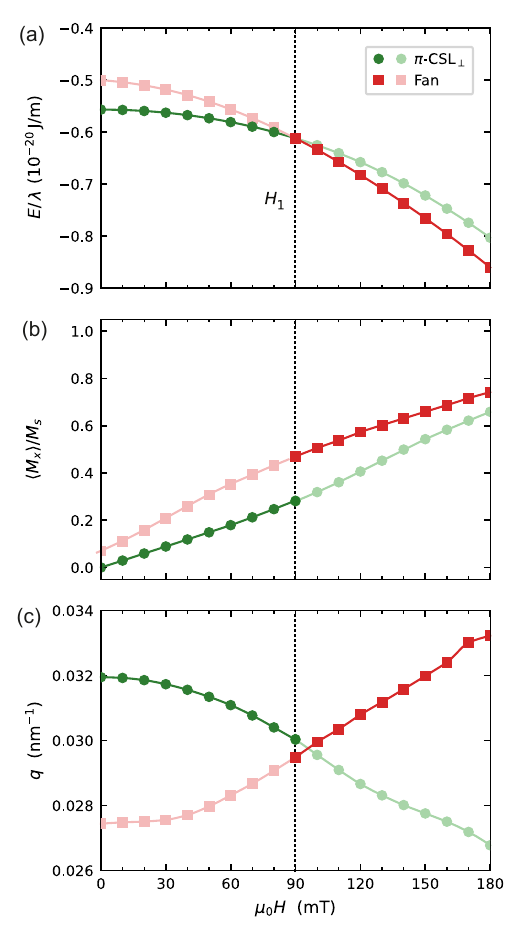}
    \caption{Field variation of 1D spin textures obtained from micromagnetic simulations of a Mn$_{1.4}$PtSn lamella with in-plane periodic boundary conditions, as shown in Fig.~\ref{fig:micro_PBCs}. Key characteristics are shown up to 180\,mT, where the slab is fully spin-polarized along the $x$-axis. (a)~Energy per modulation length relative to the spin-polarized state. The energies of the $\pi$-CSL and fan textures cross at the critical field $H_1$. Light symbols indicate the metastable (higher-energy) solution of each phase. (b)~Normalized average magnetization $\langle M_x \rangle / M_s$ along the field direction. The discontinuity between the $\pi$-CSL$_\perp$ and fan solutions at $H_1$ indicates a first-order transition. (c) The resulting equilibrium wave number $q$. At $H_1$, the $\pi$-CSL$_\perp$ solution with expanding modulation length is replaced by the fan texture, which steadily contracts and continuously connects to the saturated state.}
    \label{fig:micro_PBCs_energy}
\end{figure}

In small fields, only slight distortions appear, and the mixed-domain configuration remains preserved [cf.~\figautoref[(b)]{Fig:micromagnetic_neumann}]. However, above $H_\mathrm{1} \approx \SI{90}{\milli\tesla}$, the magnified ROIs reveal distinct responses of the two domain types: the field induces a conical distortion of the $\pi$-CSL$_\parallel$ (purple/green $\rightarrow$ pink/yellow), while in dom$_\perp$ the domain-wall orientation reverses, resulting in the formation of a fan-type spin texture (red/red). Once the field surpasses $H_\mathrm{2} \approx \SI{190}{\milli\tesla}$, domain reorganization sets in: dom$_\perp$ expands at the expense of dom$_\parallel$, while the magnified ROIs show that the conical and fan characters of the respective domains persist [cf.~\figautoref[(c)]{Fig:micromagnetic_neumann}, $\SI{200}{\milli\tesla}$]. Finally, above $H_\mathrm{3} \approx \SI{240}{\milli\tesla}$ a single-domain state is established, with a fan-type spin texture throughout the sample [cf.~\figautoref[(d)]{Fig:micromagnetic_neumann}].

\subsection{Single-domain simulations}

For deeper insight into the competition between the $\pi$-CSL and fan textures within the $\mathrm{dom}_{\perp}$ domain, we performed additional micromagnetic simulations in a single-domain configuration. The magnetization was initialized separately with a helical ($Q = 1$) and a fan ($Q = 0$) ansatz, each propagating along the $y$ axis. The system was modeled as a slab with a thickness of $100\,\mathrm{nm}$, matching the experimental lamella, while periodic boundary conditions were applied along the propagation direction. To determine the equilibrium modulation length, a magnetic field was applied along the $x$ direction, and the system size $L_y$ along the propagation direction was systematically varied. For each choice of $L_y$, the magnetization was relaxed, and the resulting energies per unit length were compared to identify the energetically preferred modulation period as a function of the applied field. This approach effectively represents an extended system and enables a reliable determination of the intrinsic modulation length of the textures. Further details of this procedure are provided in Sec.~S-II~D of the Supplemental Material.

\figautoref[(a)]{fig:micro_PBCs} shows a cross-section through the film along the propagation of the periodic texture obtained by relaxing a helical state at zero field. The bulk magnetization, indicated by arrows at $z = \SI{50}{\nano\meter}$, corresponds to a left-handed $\pi$-CSL$_\perp$ propagating along~$y$. The combination of complementary colors in the \textit{hsl} color space on adjacent solitons confirms the Bloch-type $2\pi$ rotation of the magnetization vector within each period. At the surfaces ($z = \SI{10}{\nano\meter}$, $\SI{90}{\nano\meter}$), the $\pi$-solitons are distorted towards a Néel-type rotation with magnetization components along the direction of propagation. In effect, the system imitates surface flux-closure domains that minimize the magnetostatic energy at the open boundaries, reminiscent of chiral surface twists in systems with isotropic DMI~\cite{Leonov2016,Schneider2018,Wolf2022}.

The alternative solution for the 1D spin texture in the dom$_\perp$ domain is illustrated in \figautoref[(b)]{fig:micro_PBCs}. Here, the system has been initialized in the fan state, and relaxed at $H_1<H=\SI{180}{\milli\tesla}$. Accordingly, the magnetization is predominantly aligned with the $x$ axis, which appears as varying shades of red in the \textit{hsl} color space. The spins no longer perform full rotations but merely symmetrical oscillations around the field direction. While the flux-closure distortions at the surfaces are significantly reduced in amplitude compared to the $\pi$-CSL$_\perp$ state. The comparison of the two solutions in \figautoref{fig:micro_PBCs} highlights that the equilibrium modulation length of the fan state at 180\,mT is noticeably shorter than that of the $\pi$-CSL$_\perp$ at low fields --- a trend consistent with the experimentally observed shift of the $\vect{q}_\perp$ Bragg peak position.

To quantify the phase competition between the $\pi$-CSL$_\perp$ and fan state, \figautoref[(a)]{fig:micro_PBCs_energy} shows the energy per modulation period of either solution ($E_\mathrm{min}/\lambda$). The respective values cross at $H_1=90$\,mT, where the fan structure becomes energetically more favorable than the CSL solution.

Strictly speaking, as the sample is three-dimensional, the order parameter space is $S^2$, and the transition is therefore not topologically protected. Nevertheless, as long as the DMI confines the spins to rotate within a single plane, the order parameter space is effectively reduced to $S^1$, and the transition between the $\pi$-CSL ($Q \neq 0$) and the fan ($Q = 0$) acquires a pseudo-topological character. To characterize it, we extract the average magnetization $\langle M_x \rangle / M_s$ along the field direction from the relaxed ground-state spin configurations at each field, shown in \figautoref[(b)]{fig:micro_PBCs_energy}. At low fields, $\langle M_x \rangle / M_s$ increases gradually, reflecting the continuous canting of the rotating spins toward $\vect{H}$. At $H_1$, by contrast, $\langle M_x \rangle / M_s$ jumps discontinuously. Since $\langle M_x \rangle / M_s$ is the thermodynamic variable conjugate to the applied field, i.e., a first derivative of the free energy, this discontinuity classifies the transition as first order in the Ehrenfest sense. The discontinuity reflects the fact that the $\pi$-CSL cannot be deformed globally and continuously into the fan while shedding its net winding. Microscopically, the transition instead proceeds through a lateral unwinding of the $\pi$-solitons. A video file provided with the Supplementary Material illustrates how this unwinding nucleates at imperfections of the magnetic texture and rapidly advances along the domain walls.

We stress that this classification rests on the simulated order parameter; no thermodynamic quantities have been measured in the present experiment. To confirm the character of the transition, bulk magnetometry or calorimetry across $H_1$ would be highly desirable, but are hardly feasible for nanoscale crystals. On the other hand, we note that even subtle details of the present REXS data are qualitatively consistent with the simulations. For instance, the reduction of the dom$_\perp$ Bragg-peak intensity at $H_1$ [cf.~\figautoref[(b)]{fig_fieldsweep_0_to_sat_bragg_only_magnitude}] matches the loss of the out-of-plane magnetization component, and the abrupt increase of $\langle M_x \rangle / M_s$ expected upon the collapse of the $\pi$-solitons.

The corresponding field evolution of the equilibrium wave number $q$ of either spin texture is shown in \figautoref[(c)]{fig:micro_PBCs_energy}. In the $\pi$-CSL$_\perp$ state (green markers), $q$ decreases monotonically with increasing field, reflecting the growth of the modulation length as the spacing between solitons expands --- the expected behavior of a chiral soliton lattice approaching its commensurate--incommensurate transition. The fan state (red) exhibits the opposite trend: $q$ grows proportionally to the applied field, corresponding to a shrinking modulation length as the fan amplitude diminishes toward the field-polarized state.

The first-order transition and the steadily contracting fan texture above $H_1$ agree well with the experimental observations. Below $H_1$, however, the expansion of the $\pi$-CSL$_\perp$ modulation predicted by the single-domain simulations is not observed in the REXS experiment; instead, the Bragg angle remains approximately constant up to the transition. This discrepancy stems from the idealized nature of these simulations, which describe an isolated, perfectly periodic 1D texture free to adjust its modulation length, and thus cannot capture the mixed-domain character of the real sample.

As seen in our multi-domain simulations and reported in previous studies~\cite{Nayak2017,Peng2020,Jena2020a,Ma2020,ZunigaCespedes2021}, the stripes of adjacent orthogonal domains connect continuously across the domain boundaries, thereby coupling the modulation lengths of the two domains. Below $H_1$, the domains respond to an in-plane field with opposite tendencies: the conically distorted $\pi$-CSL$_\parallel$ favors a contracting period, whereas the $\pi$-CSL$_\perp$ favors an expanding one. Their mutual connection may thus frustrate both responses and lock the common modulation length, consistent with the observed plateau of $q_\parallel$ and $q_\perp$.

This interpretation is corroborated by the complementary case of out-of-plane fields~\cite{Winter2025}: a field along the $c$ axis respects the tetragonal symmetry and acts equivalently on both domains, which accordingly respond in unison and exhibit the CSL-like expansion of the modulation period. The locking observed here is therefore not a generic suppression of period changes, but a consequence of the inequivalent response of the two domains to the symmetry-breaking in-plane field. Once $H_1$ is crossed, both the fan and the conical texture favor a contracting period; the competition is lifted, and the modulation lengths of both domains decrease in unison --- precisely the behavior observed in the experiment.

\section{Discussion and Conclusion}

In this work, we have investigated this scenario using a non-centrosymmetric Heusler compound with $D_{\mathrm{2d}}$ point symmetry and anisotropic DMI.

While spin textures in Mn$_{1.4}$PtSn have been thoroughly studied using LTEM~\cite{Nayak2017,Peng2020,Jena2020a,Ma2020}, the response under applied in-plane fields has largely remained experimentally inaccessible, as the technique is incompatible with fields applied perpendicular to the electron beam. Resonant elastic X-ray scattering combined with a vector magnet proves to be an invaluable complementary tool, enabling quantitative characterization of the Fourier components of spin textures in momentum space and direct comparison with micromagnetic simulations.

The presence of a two-domain CSL state at zero field allowed tracking the behavior with fields parallel and perpendicular to the propagation vector in the same sample. For the CSL propagating perpendicular to the field, we evidenced a first-order phase transition at a critical field $H_1$. Here, the chiral solitons abruptly unwind to a fan texture, which is continuously connected to the spin-polarized state. The other domain, whose propagation direction is already aligned with the field, avoids transition to a conical state as it is rapidly displaced by the ``perpendicular'' domain between two characteristic fields $H_2$ and $H_3$.

A micromagnetic model of this scenario allowed us to corroborate both the topological--like transition at $H_1$, and the domain redistribution at $H_2\sim H_3$. Moreover, using periodic boundary conditions in the plane of the sample and energetically optimizing the modulation length, we were able to reproduce the experimentally observed contraction of the fan state proportional to the applied field.

Crucially, while the DMI penalizes the fan configuration in the bulk --- as the symmetric oscillation of spins about the field direction is incompatible with a single rotational sense --- our simulations reveal that magnetostatic interactions in finite-thickness films provide the additional energy gain required to stabilize it as the true ground state above $H_1$. {The surface flux-closure domains highlight the important role of magnetostatic interactions in shaping the equilibrium spin texture. In
particular, they can stabilize fan-like structures even in the presence of the DMI, which favors a single rotational sense.

These results complete a unified picture of field-driven transformations of one-dimensional spin textures in anisotropic chiral magnets. As previously reported, out-of-plane fields condense $\pi$-solitons into higher-order $2\pi$-CSLs that evolve continuously toward the ferromagnetic state, preserving their topological character throughout~\cite{Winter2025}. By contrast, in-plane fields drive a discontinuous unwinding into the topologically trivial fan. This dual tunability --- continuous versus discontinuous, topology-preserving versus topology-destroying --- showcases how field orientation can act as a switch between fundamentally different transformation pathways in chiral magnets.

Notably, the mechanism that we have here identified --- the magnetostatic stabilization of a nonchiral modulated state in a finite film of a chiral magnet --- is not specific to the $D_{2d}$ symmetry of Mn$_{1.4}$PtSn. It should apply equally to other symmetry classes hosting one-dimensional modulations, including $C_{nv}$ helimagnets such as the FeGe~\cite{Ukleev2020} and CrNb$_3$S$_6$~\cite{Togawa2012} families, where analogous fan instabilities may have gone undetected. The ability to toggle irreversibly between topologically distinct one-dimensional states using moderate in-plane fields opens pathways for field-programmable control of spin-polarized state topology, with potential applications in non-volatile spintronic and magnonic devices where the topological invariant itself encodes information.

\begin{acknowledgments}
The authors thank J.\ Masell for valuable discussions on the topic. M.Wi.\ acknowledges support from the International Max Planck Research School for Chemistry and Physics of Quantum Materials (IMPRS-CPQM). M.C.R.\ was supported through the Emmy Noether programme of the German Science Foundation DFG (project no.\ 501391385). Work at TU Dresden was supported by the DFG through CRC 1143 and the Cluster of Excellence ct.qmat (EXC 2147, project ID 390858490). K.E.S. acknowledges funding from the DFG through the SPP Skyrmionics through project no.\ 403233384 and project no.\ 505561633 in the TOROID project co-funded by the French National Research Agency (ANR) under Contract No.\ ANR-22-CE92-0032. M.Wi, A.T., and B.R.\ are grateful for funding from the DFG vie the SPP 2137, project no.\ 403503416. M.A.\ acknowledges funding from the UDE Postdoctoral Seed Funding program. Experiments were performed on the Portable Octupole Magnet System on beamline I10 at the Diamond Light Source, UK, under proposal MM28882, at  VEKMAG at BESSY-II (Germany), under proposal 231-11833-ST. Financial support by the EPSRC (EP/N032128/1) is gratefully acknowledged. The STXM experiment was carried out at the BESSY II synchrotron as part of the proposal Photons-251-13224. We thank the Helmholtz-Zentrum Berlin f\"ur Materialien und Energie for the allocation of beamtime. V.U., C.L., F.R. acknowledge financial support for the VEKMAG project and for the PM2-VEKMAG beamline by the German Federal Ministry for Education and Research (BMBF 05K2010, 05K2013, 05K2016, 05K2019) and by HZB. Parts of the manuscript were stylistically optimized with the aid of a large language model. The authors reviewed, revised, and take full responsibility for the final manuscript.
\end{acknowledgments}

\section*{Author Contributions}
M.Wi., M.C.R., J.G., C.F.,\ and B.R.\ devised the project.
M.Wi., M.C.R., A.S.S., B.A., J.R.B., V.U., C.L., F.R., S.W., M.We., A.M., G.vdL., and T.H. conducted the REXS and STXM experiments. M.Wi. conducted the multi-domain micromagnetic simulations. A.P., M.A., and K.E.S.\ developed the theory and conducted single-domain micromagnetic simulations in close collaboration with M.Wi.
M.Wi.\ analyzed all data.
M.Wi., A.T., and P.V.\ prepared the samples.
M.W., B.R., M.C.R., T.H., A.P., K.E.S., and M.A.\ wrote the manuscript. All authors contributed to the discussion and reviewed the manuscript.


\bibliographystyle{apsrev4-2}
\bibliography{mps_arxiv}

\clearpage
\onecolumngrid


\section*{Supplementary material}

# Supplemental Material for
# Quasi-1D Spin Textures: From Chiral Soliton Lattice to Fan State

M. Winter, A. Pignedoli, A. S. Sukhanov, M. Azhar, A. Tahn, B. Achinuq, J. R. Bollard, V. Ukleev, C. Luo, F. Radu, S. Wintz, M. Weigand, A. Mistonov, P. Vir, J. Geck, C. Felser, G. van der Laan, T. Hesjedal, K. Everschor-Sitte, B. Rellinghaus, and M. C. Rahn



## S I. REXS PATTERNS IN DECREASING IN-PLANE FIELDS

The field was ramped to 450 mT to ensure full saturation and subsequently reduced to zero. The corresponding REXS patterns at 0, 150, 250, and 350 mT are shown in Figs. S1 (a–d). Only the Bragg peaks at $\boldsymbol{q}_\perp$ reappear. The sequence demonstrates that the field-induced collapse of the domain propagating along the field direction is irreversible, leaving the sample in a single-domain state down to zero field. Furthermore, the position of the $\boldsymbol{q}_\perp$ Bragg peak shifts continuously to smaller momentum transfer as the field is reduced, i.e., the modulation length expands again upon decreasing field. This behavior mirrors, in reverse, the continuous contraction of the modulation observed during the up-sweep [cf. Fig. 3(c) of the main text] and confirms that the modulation length of the fan texture is a reversible, single-valued function of the applied field.

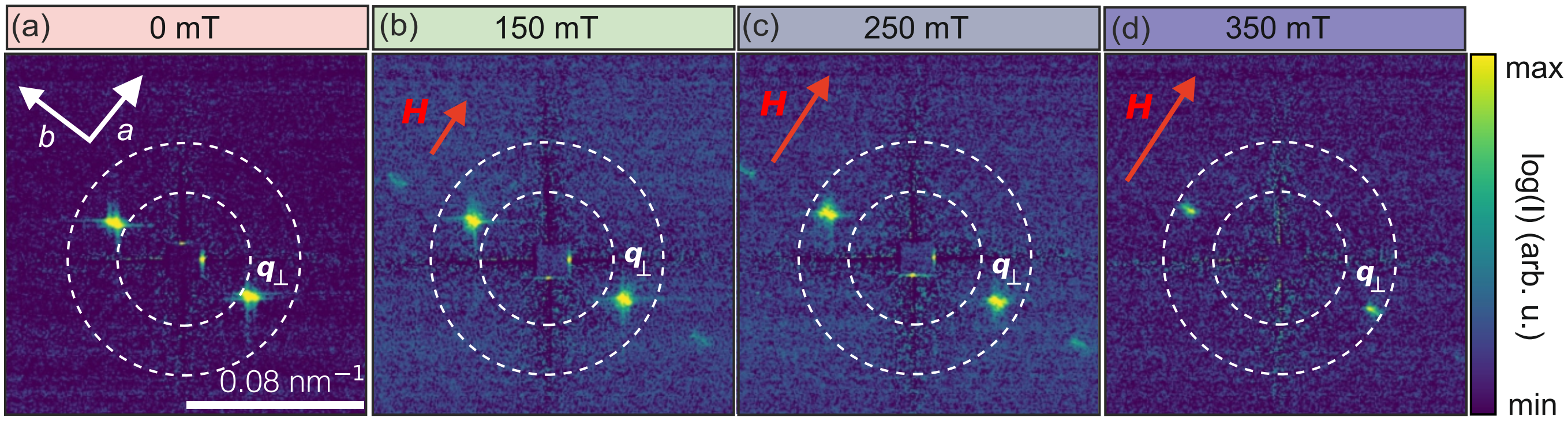


FIG. S1. REXS patterns measured under decreasing in-plane fields applied parallel to the $a$-axis. The background-subtracted intensity is shown on a logarithmic color scale with an additional median filter. Panels (a–d) display the single-domain state at 0, 150, 250, and 350 mT, respectively.

## S II. SCANNING TRANSMISSION X-RAY MICROSCOPY

Real-space investigation of the in-plane field-driven domain reorganization was carried out by means of scanning transmission X-ray microscopy (STXM) using the MAXYMUS instrument at the BESSY II synchrotron (Helmholtz-Zentrum Berlin, Germany). Measurements were performed at room temperature, with in-plane magnetic fields of 0–250 mT provided by a system of permanent magnets [1]. A circularly polarized soft X-ray beam was tuned to the Mn $L_2$ absorption edge ($E = 653\,\mathrm{eV}$) to optimize the balance between absorption and X-ray magnetic circular dichroism (XMCD) contrast.

Note that these measurements were performed on a similar but distinct lamella, whose critical fields may differ from the values $H_1$–$H_3$ determined by REXS and are difficult to determine from real-space imaging alone — REXS is the more suitable probe in this respect. The STXM data should therefore be regarded as a qualitative, real-space confirmation of the domain reorganization process.

STXM images are shown in Fig. S2. At zero field, a two-domain $\pi$-CSL state is clearly visible [Fig. S2(a)], and the corresponding fast Fourier transform (FFT) patterns show four distinct spots, consistent with the REXS data. The XMCD contrast decreases as the in-plane field is increased to 180 mT and 250 mT [Fig. S2(b,c)], reflecting the canting of the magnetic moments toward the field direction. While no obvious changes in the stripe pattern are observed at 180 mT [Fig. S2(b)], at 250 mT [Fig. S2(c)] the energetically unfavorable $\boldsymbol{q}_\parallel$ domain develops directional disorder, and its wave vector partially reorients perpendicular to the field axis. The corresponding FFT pattern shows two broad

maxima along that direction, in addition to a diffuse ring that is also noticeable in the REXS data [cf. Fig. 2(c) of the main text].

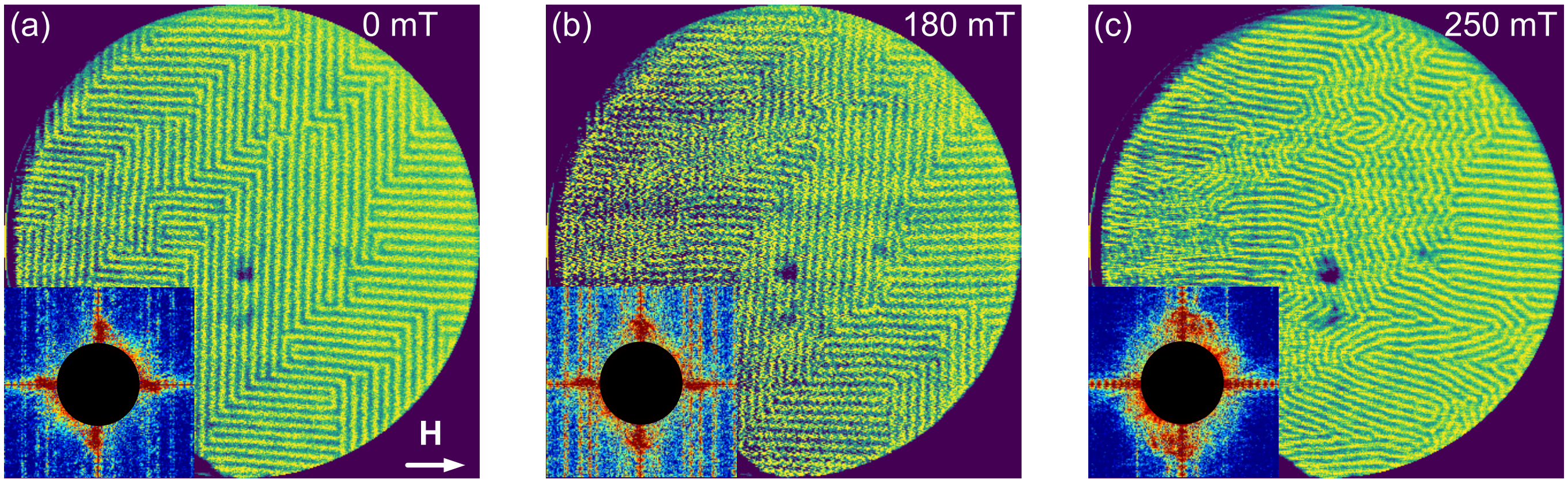


FIG. S2. STXM images of the field-driven domain reorganization as a function of the in-plane magnetic field: (a) 0 mT, (b) 180 mT, and (c) 250 mT. Insets show the corresponding FFT patterns. The measurements were obtained on a different lamella than the REXS data of the main text.

## S III. COMPUTATIONAL METHODS

### A. Micromagnetic Hamiltonian for $Mn_{1.4}PtSn$ under $D_{2d}$ Symmetry

The non-centrosymmetric Heusler compound $Mn_{1.4}PtSn$ with $D_{2d}$ point symmetry can be modeled by the micromagnetic Hamiltonian

$$\mathcal{H} = \int d\boldsymbol{r}\,[\varepsilon_{\rm ex} + \varepsilon_{\rm DMI} + \varepsilon_{\rm Z} + \varepsilon_{\rm MCA} + \varepsilon_{\rm MS}]\,, \tag{S1}$$

which includes ferromagnetic exchange, anisotropic DMI [2–4], Zeeman coupling, magnetocrystalline anisotropy (MCA), and magnetostatic (MS) contributions:

$$\begin{aligned}
\varepsilon_{\rm ex} &= A\,(\boldsymbol{\nabla}\boldsymbol{m})^2\,,\\
\varepsilon_{\rm DMI} &= D(m_z\partial_x m_y - m_y\partial_x m_z - m_x\partial_y m_z + m_z\partial_y m_x)\,,\\
\varepsilon_{\rm Z} &= -\mu_0\,M_s\,\boldsymbol{m}\cdot\boldsymbol{H}_{\rm ext}\,,\\
\varepsilon_{\rm MCA} &= -K_{\rm MCA}\,(\boldsymbol{m}\cdot\hat{\boldsymbol{n}})^2\,,\\
\varepsilon_{\rm MS} &= -\mu_0\,M_s\,(\boldsymbol{m}\cdot\boldsymbol{H}_{\rm MS})/2\,.
\end{aligned} \tag{S2}$$

Here, $\boldsymbol{m}(\boldsymbol{r})$ denotes the unit magnetization vector. The exchange stiffness $A$ favors collinear spin alignment, $D$ is the DMI constant, $M_s$ is the saturation magnetization, and $K_{\rm MCA}$ is the uniaxial anisotropy constant. The magnetostatic field $\boldsymbol{H}_{\rm MS}$ is computed self-consistently by the solver. The temporal evolution of the magnetization is governed by the Landau–Lifshitz–Gilbert (LLG) equation, including both precession and damping.

### B. Material Parameters

The material parameters for $Mn_{1.4}PtSn$ were taken from Ref. [5]. The exchange stiffness is $A = 8\times10^{-12}$ J/m, the DMI constant $D = 0.4\times10^{-3}$ J/m$^2$, and the saturation magnetization $M_{\rm s} = 4.45\times10^5$ A/m. The uniaxial magnetocrystalline anisotropy is $K_{\rm MCA} = 1\times10^5$ J/m$^3$, with the easy axis $\hat{\boldsymbol{n}}$ along the crystallographic $c$ axis. For dynamical simulations, a Gilbert damping $\alpha = 0.03$ and gyromagnetic ratio $\gamma = 2.211\times10^5$ m/(A s) were used.

### C. Finite-sample simulations with open boundaries

Micromagnetic simulations were performed using `OOMMF` [6] via the Python interface `Ubermag` [7]. The system was modeled as a finite sample of size $2\,\mu\text{m} \times 2\,\mu\text{m} \times 100\,\text{nm}$, discretized with a mesh of $10\,\text{nm} \times 10\,\text{nm} \times 10\,\text{nm}$. Neumann boundary conditions were applied in all directions. All simulations were initialized from the same magnetization state $\boldsymbol{m}_0$. To reproduce the experimental ground state of $Mn_{1.4}PtSn$, the initial configuration included spiral phases with wavelengths consistent with experiment. A mixed-domain state was constructed by dividing the sample into four regions in the $xy$-plane, each initialized with spirals of alternating chirality and orthogonal propagation directions. This resulted in a multidomain configuration at zero field, consistent with the experimental starting conditions. The system was first relaxed at zero field to minimize the total energy. The external field was then increased in steps of $10\,\text{mT}$ up to $450\,\text{mT}$ along the $x$ direction, using the relaxed state from the previous field step as the initial condition.

### D. Spiral and fan states with periodic boundary conditions

The finite-sample simulations described above are well suited to capture domain evolution and boundary-mediated processes, but they are not appropriate for determining the equilibrium modulation length of spiral and fan textures: because the system size is fixed, the modulation period is constrained by the simulation geometry, and additional energy barriers can arise at the sample boundaries. As a result, these simulations do not reliably reflect the intrinsic field dependence of the texture wavelength. To overcome this limitation, we performed complementary micromagnetic simulations using periodic boundary conditions (PBCs) along the propagation direction of the texture. This approach removes boundary-induced constraints and allows the system to represent an effectively extended medium. For each applied magnetic field, we varied the system length $L_y$ along the propagation direction and relaxed the magnetization for each choice of $L_y$. By comparing the resulting energies per unit length, we identified the energetically preferred modulation period for both types of textures. This procedure provides direct access to the intrinsic field dependence of the wave vector and enables a meaningful comparison of the energetics of spiral and fan states, free from the artificial constraints imposed by finite-size simulations.

By using PBCs and appropriate initializations, the simulations can directly capture the topological distinction between the spiral and fan phases. To this end, the system was initialized separately with a helical ansatz ($Q = 1$) and a fan ($Q = 0$) ansatz, each propagating along the $y$ axis. The helical ansatz, Eq. (S3), corresponds to a left-handed Bloch helix,

$$\begin{aligned} m_x &= \sin\left(\frac{2\pi}{L_y} y\right), \\ m_y &= 0, \\ m_z &= \cos\left(\frac{2\pi}{L_y} y\right), \end{aligned} \tag{S3}$$

while the fan ansatz, Eq. (S4), describes a symmetric oscillation of the magnetization about the field ($x$) direction,

$$\begin{aligned} m_x &= \sin\left(\frac{\pi}{3} \cos\left(\frac{2\pi}{L_y} y\right) + \frac{\pi}{2}\right), \\ m_y &= 0, \\ m_z &= \cos\left(\frac{\pi}{3} \cos\left(\frac{2\pi}{L_y} y\right) + \frac{\pi}{2}\right). \end{aligned} \tag{S4}$$

In both cases, $m_x$, $m_y$, and $m_z$ are the components of the unit magnetization. The energetically preferred period $L_y$ was determined by systematically varying $L_y$ at each specified magnetic field. The configuration that achieved the minimal energy per unit length, $E_{\min,L}/L_y$, was selected as the equilibrium period for the given field. To incorporate surface-related effects, the slab thickness was set to match the experimental sample thickness, $L_z = 100\,\text{nm}$. The mesh cell size were $v_{\text{mesh}} = 1\,\text{nm} \times 1\,\text{nm} \times 1\,\text{nm}$. Given the one-dimensional, single-$q$ ansatz, the simulation cell was restricted to a single mesh unit along the $x$ direction, $L_x = 1\,\text{nm}$.

By repeating this procedure for different field strengths, the field-dependent evolution of the equilibrium periods $L_{\text{sim}}^{\text{Bloch}}(H)$ and $L_{\text{sim}}^{\text{fan}}(H)$, as well as the corresponding energy minima $E_{\min,L}^{\text{Bloch}}(H)$ and $E_{\min,L}^{\text{fan}}(H)$ of the spiral ($Q = 1$) and fan ($Q = 0$) states, respectively, were determined.



\end{document}